# HD and $H_2$ formation in low-metallicity dusty gas clouds at high redshift

S. Cazaux[1], M. Spaans[1]

Kapteyn Astronomical Institute, PO box 800, 9700 AV Groningen, The Netherlands

**ABSTRACT**

*Context.* The HD and $H_2$ molecules play important roles in the cooling of primordial and very metal-poor gas at high redshift.

*Aims.* Grain surface and gas phase formation of HD and $H_2$ is investigated to assess the importance of trace amounts of dust, $10^{-5} - 10^{-3} Z_\odot$, in the production of HD and $H_2$.

*Methods.* We consider carbonaceous and silicate grains and include both physisorption and chemisorption, tunneling, and realistic grain surface barriers. We find, for a collapsing gas cloud environment with coupled chemical and thermal balance, that dust abundances as small as $10^{-5}$ solar lead to a strong boost in the $H_2$ formation rate due to surface reactions. As a result of this enhancement in $H_2$, HD is formed more efficiently in the gas phase through the $D^+ + H_2$ reaction. Direct formation of HD on dust grains cannot compete well with this gas phase process for dust temperatures below 150 K. We also derive up-to-date analytic fitting formulae for the grain surface formation of $H_2$ and HD, including the different binding energies of H and D.

*Results.* Grain surface reactions are crucial to the availability of $H_2$ and HD in very metal-poor environments. Above metallicities of $10^{-5}$ solar, the grain surface route dominates the formation of $H_2$, which in turn, drives the formation of HD in the gas phase. At dust temperatures above 150 K, laboratory experiments and theoretical modelling suggest that $H_2$ formation on grains is suppressed while HD formation on grains is not.

## 1. Introduction

The chemistry occurring in primordial gas in the early universe involves mainly hydrogen, deuterium, helium and lithium, as well as their ionic forms (Glover & Abel 2008; Ripamonti 2007; Yoshida et al. 2006; Galli & Palla 1998). The formation of the first stars crucially depends on the availability of specific molecular coolants like $H_2$ and HD. The molecule $H_2$ allows gas to cool down to a few hundred Kelvin, because its first rotationally excited state lies at about 500 K. The molecule HD, which unlike $H_2$ possesses a small dipole moment, allows cooling to below 100 K, given that its first excited state lies at about 150 K. Simulations that incorporate non-equilibrium chemistry and cooling of primordial gas in the early universe (e.g., Abel et al. 2000, Bromm et al. 2002, Abel et al. 2007, Wise et al. 2007; 2008, Klessen et al. 2007, Jappsen et al. 2007) find that

*Send offprint requests to*: cazaux@astro.rug.nl

these molecules are instrumental to the collapse of clouds in young galaxies and the formation of the first stars. Once the first stars were formed and ionised the universe, the next generation of stars could be formed from more ionized gas. It is generally thought that the final masses of the first stars are a few dozen to a hundred solar masses (Johnson & Bromm 2006, Yoshida et al. 2007). In this, an increased electron abundance boosts the formation of $H_2$, and lowers the gas temperature to below 200 K, relative to less ionized gas. Subsequently, HD cooling can become more important than $H_2$. Recent work by McGreer & Bryan (2008) on zero metallicity gas shows that HD cooling is dominant for halos with masses below $10^{5.5}$ $M_\odot$, yielding stars that are 6 times less massive. Also, HD cooling in ionized halos is most effective for a density range between $10^2$-$10^6$ cm$^{-3}$, while above this range $H_2$ cooling dominates again. In all, it is of great importance to determine the chemical composition of collapsing clouds with density, in order to establish which coolant ($H_2$, HD, other) dominates, and which mass of star results from gravitational collapse.

The molecules $H_2$ and HD can be formed in the gas phase at zero metallicity. For HD, the dominant reactions are $D^+ + H_2 \rightarrow HD + H^+$ (exothermic) for its formation, and $H^+ + HD \rightarrow H_2 + D^+$ (-962K) for its destruction (Glover & Abel 2008). For $H_2$, one has $H^- + H \rightarrow H_2 + e$. Additional relevant reactions are $H^- + D \rightarrow HD + e$; $D^- + H \rightarrow HD + e$ and $D^+ + H_2 \rightarrow HD + H^+$. These reactions typically proceed during free-fall collapse, i.e., in a time dependent environment. Since the gas starts out at temperatures of about $10^3$ K or more prior to collapse, HD destruction is efficient. As the $H_2$ abundance increases with time, so does the cooling rate and the temperature (as well as the $H^+$ abundance) drops, stimulating the presence of HD. However, once some pollution by metals has occurred, i.e., after the very first stars have exploded as supernovae (SNe), dust grains can be present as well in the ambient interstellar medium (ISM), see work by Todini & Ferrara (2001) and Bianchi & Schneider (2007). The presence of dust at high redshift, as observed toward a QSO at z~6.2 (Maiolino et al. 2004), requires efficient condensation of grains in SN ejecta. Models of dust formation in ejecta of SNe (Bianchi & Schneider 2007, Todini & Ferrara 2001) present the grain size distribution of silicates, amorphous carbon (AC), magnetite and corundum; and show that the largest grains are the AC ones, with sizes around 300Å, whereas the other grain types have smaller radii, around 10-20Å. Once the first grains are produced by SNe, they will influence the next generation of stars as some species will start to form on dust grains. Consequently, additional pathways for $H_2$ and HD formation open up.

Efforts that aim to include both gas and grain surface reactions through an equation of state analysis can be found in Omukai et al. (2005) and Spaans & Silk (2005). In this, the pressure $P$ is usually assumed to follow the polytropic form $P \sim \rho^\gamma$, for the mass density $\rho$ and polytropic exponent $\gamma$. More detailed hydrodynamical simulations of gas phase chemistry only can be found in Smith et al. (2008), while first steps to include some aspects of grain chemistry can be found in Glover & Jappsen (2007). All these works show that the interplay between gravity and thermodynamics acts to a large degree through the amount of fragmentation, through the Jeans mass, that a collapsing gas cloud experiences, thus setting the typical masses of stars. In this, the occurrence of fragmentation follows from whether the gas temperature rises or decreases under compression. I.e., whether $\gamma$ is larger of smaller than unity. This change in temperature under compression can be a strong function of density for $0.1 < n < 10^{17}$ cm$^{-3}$, and depends on the ambient metallicity (Omukai et al. 2005; Spaans & Silk 2005). The aim of this work is then to investigate the importance of HD and $H_2$ formation on grain surfaces and their influence on gas thermodynamics, and to provide analytic fits for $H_2$ and HD formation on grains that are easy to implement in cosmological simulations of early structure formation.

## 2. Grain surface chemistry

### 2.1. Formation efficiencies of $H_2$ and HD on dust surfaces

In a previous paper (Cazaux et al. 2008), we discussed the formation of $H_2$ and HD on surfaces that are typical of the ISM. We found that the formation of molecules depends on the binding energy of atoms with the surface and on the barrier that atoms from the gas phase have to cross in order to become strongly bound to the surface. Indeed, there are two interactions between the atoms and the surface: a weak one, called physisorption (Van der Waals interaction), and a strong one, called chemisorption (covalent bond), as represented in fig. 1. Atoms on the grain surface can move from site to site by tunnelling effects and thermal hopping. Atoms from the gas phase can access easily the physisorbed sites and become physisorbed atoms. These weakly bound atoms can scout the surface at very low dust temperatures, and can meet each other to form molecules. Once the dust temperature becomes higher, the physisorbed atoms evaporate and the formation of molecules is insured by the contribution of strongly bound (chemisorbed) atoms. Depending on the magnitude of the barrier that needs to be crossed to access the chemisorbed sites, a fraction of physisorbed atoms can enter the chemisorbed sites and meet an already chemisorbed atom to form molecules, but, if the barrier is very high, atoms from the gas phase, which have higher energy, cross the barrier to enter directly into chemisorbed sites and form molecules. These processes allow molecules to form for a wide range of dust grain temperatures (Cazaux et al. 2008, Cazaux & Tielens 2002, Cazaux & Spaans 2004).

We have developed a rate equations method to describe the chemistry occurring on interstellar dust grains. This method follows the populations of the different species on the grain (physisorbed H, D, $H_2$ and HD and chemisorbed H and D). The different processes that can occur in this model are the following: 1) atoms from the gas phase accrete into a physisorbed or chemisorbed site; 2) physisorbed atoms go to another physisorbed site, or to a chemisorbed site; 3) chemisorbed atoms evaporate. The mechanisms to form molecules are either through the Langmuir-Hinshelwood kinetic (an atom on the surface moves into an already occupied site) or the Eley-Rideal kinetic (an atom from the gas phase arrives in an occupied site). With this method, we obtain the efficiencies of the formation of $H_2$ and HD on interstellar dust grains (note that $D_2$ has been treated in Cazaux et al. 2008 and is not studied here). These efficiencies, as function of dust grain temperature, are reported in figure 2. The formation of $H_2$ and HD is very efficient on all types of grains at low temperatures ($\leq 20$ K), because it involves physisorbed atoms. At higher dust temperatures, the chemisorbed atoms become relevant. Depending on the barrier against chemisorption, the physisorbed atoms may be able to enter chemisorbed sites, as is the case for an amorphous carbon surface (no barrier), or the physisorbed atoms evaporate before entering the chemisorbed sites, as is the case for silicates and graphitic surfaces (high barrier). In the latter case, the formation of molecules is insured by atoms that come from the gas phase and enter directly into chemisorbed sites. Consequently, for intermediate dust temperatures ($\geq 20$K), the efficiency decreases as the barrier against chemisorption increases.

In the ISM, dust grains are mainly carbonaceous particles or silicates, with various sizes, and a large fraction of the available surface for chemistry is in the form of very small grains or polycyclic aromatic hydrocarbons (PAHs; Weintgarner & Draine 2001). Small carbon grains exist mostly in the form of PAHs, while big carbon grains occur as amorphous carbon. These two types of grains have different surface properties, as discussed in Cazaux et al. (2008). PAHs have surfaces similar to graphite, that consist in hexagonally arranged carbon atoms. Atoms that are coming on these surfaces can be placed in these hexagonal structures in different configurations. Ortho refers to two neighboring atoms, meta refers to two atoms which are separated, but not opposed and para refers to two opposed atoms. Recent discoveries show that the properties of PAHs and graphite surfaces depend on the presence of atoms on the surfaces. Indeed, once an atom becomes chemisorbed on the surface, it has to cross an important barrier of 0.2 eV (Hornekær et al. 2006, Sha & Jackson 2002, Jeloaica & Sidis 1999). A second atom can become chemisorbed in a para-site without a barrier (Hornekær et al. 2006, Rougeau et al. 2006) and a third atom will form a molecule without a barrier (Bachellerie et al. 2007). Because the formation of molecules on PAHs cannot be described by rate equations, since the surface characteristics change when atoms are present on the surface, we developed a Monte Carlo method to follow the formation of $H_2$ and HD on these very small grains. Also, for silicates and amorphous carbon grains, which are bigger grains, this Monte Carlo method can be applied and compared to rate equations methods. In our model, we represent the grain by a square grid, with at each point a physisorbed or a chemisorbed site. A list of events is first calculated to determine the accretion of H and D on the grain. These times depend on the flux of H or D that arrives on the grain, and we select randomly a time for each event, following a Poisson distribution. Then, once the atoms have arrived on the grain, the list is updated taking into account the different events that can occur.

The results of our Monte Carlo simulations for graphitic surfaces (PAHs), with the inclusion of the para-site properties, are reported in fig 3, right panel. The efficiencies of $H_2$ and HD are enhanced by a few orders of magnitude in comparison to a graphitic surface that does not change its properties in the presence of atoms on its surface. We consider here a very small grid of 30 Å length, and also present, for comparison, the efficiencies of $H_2$ and HD on amorphous carbon and silicates. For these surfaces, the results are similar to the ones computed with rate equations. This is due to the fact that, as discussed in Cazaux et al. (2008), HD formation changes with grain size only for high D/H ratios, and only at very low temperatures, when physisorbed atoms are involved in the formation of molecules. Only $D_2$ formation is very sensitive to grain size changes. So because we focus our study here on the formation of $H_2$ and HD, the changes in efficiency with grain size are not important as long as the D/H ratio remains small. As a general conclusion, the $H_2$ and HD formation efficiencies are similar on graphitic surfaces (with para-site properties) and on amorphous carbon grains. The efficiencies on silicate surfaces, on the other hand, are very different because of the important barrier against chemisorption. The efficiencies for the formation of molecule are the rate of molecules that form on the surface devided by the incoming flux of atoms. The flux of incoming atoms depends on the velocity of the atoms in the gas phase $v_{at}$, the sticking coefficient $S(T_g, T_d)$, the number density of the atoms $n_{at}$, and the cross section of the dust grain as $Flux_{at} = n_{at}v_{at}\sigma S(T_g, T_d)$, where (at) can be H or D. The efficiencies for the formation of $H_2$ and HD are presented in fig 3 and can be written as follows:

$$\epsilon_{H_2} = \left(\frac{2\alpha_{ppH}H_P^2 + 2\alpha_{pcH}H_PH_C + Flux_HT_HH_C}{Flux_H}\right) \quad (1)$$

$$\epsilon_{HD} = \left(\frac{\alpha_{ppH}H_PD_P + \alpha_{pcH}H_PD_C + \alpha_{pcD}D_PH_C + Flux_DT_DH_C}{Flux_D}\right) \quad (2)$$

These equations show the principal mechanisms for the formation of molecules at different grain temperatures: 1) At low surface temperatures Td, an H (D) atom moves to a filled physisorbed site with a mobility $\alpha_{ppH}$ ($\alpha_{ppD}$). 2) When physisorbed atoms start to evaporate, with a rate $\beta_{H_P}$ ($\beta_{D_P}$), some of them enter the chemisorbed site, with a mobility $\alpha_{pcH}$ ($\alpha_{pcD}$) and meet a chemisorbed atom. 3) At higher temperatures, atoms coming from the gas phase with a temperature Tg can directly enter a chemisorbed site and form a molecule. They need to pass the barrier against chemisorption with a probability $T_H$ for hydrogen, and $T_D$ for deuterium. The probabilities and mobilities to go from one site (physisorbed or chemisorbed) to another site are determined by the transmission coefficients to cross the barrier that separates these two sites (for details see Cazaux & Tielens 2004).

Under steady state conditions, $H_P = \frac{Flux_H(1-T_H)}{\alpha_{pcH}+\beta_{H_P}}$, $H_C = \frac{1}{2}$, $D_P = \frac{Flux_D(1-T_D)}{\alpha_{pcD}+\beta_{D_P}}$ and $D_C = \frac{\alpha_{pcD}D_P}{2\alpha_{pcH}H_P}$.

The formation of $H_2$ and HD, due to the association of physisorbed and chemisorbed atoms at low surface temperature, and to the association of gas phase atoms arriving in chemisorbed sites at high surface temperature, can be approximated as:

$$\epsilon_{H_2} = \frac{2}{Flux_H}\left(\alpha_{pcH}H_PH_C\right) + \frac{1}{Flux_H}Flux_HT_HH_C = \left(\frac{\alpha_{pcH}(1-T_H)}{\alpha_{pcH}+\beta_{H_P}} + \frac{T_H}{2}\right) \quad (3)$$

$$\epsilon_{HD} = \frac{1}{Flux_D}\left(\alpha_{pcH}H_PD_C + \alpha_{pcD}D_PH_C\right) + \frac{1}{Flux_D}Flux_DT_DH_C = \left(\frac{\alpha_{pcD}(1-T_D)}{\alpha_{pcD}+\beta_{D_P}} + \frac{T_D}{2}\right) \quad (4)$$

The different mobilities of the atoms to go from a physisorbed site to a chemisorbed site (see fig 1 for the meaning of the different parameters) can be approximated as follows:

$$\alpha_{pcH} = \begin{cases} \frac{16\nu_{H_P}T_d}{(E_{chem}-E_S)}\exp\left(-2a_{pc}\sqrt{\frac{2m_Hk(E_{phys}-E_S)}{\hbar^2}}\right) = \frac{16\nu_{H_P}T_d}{(E_{chem}-E_S)}\exp\left(-4\times 10^9 a_{pc}\sqrt{(E_{phys}-E_S)}\right) & \text{; if } E_S < 0 \\ 4\nu_{H_P}\left(1+\sqrt{\frac{E_{chem}-E_S}{E_{phys}-E_S}}\right)^{-2}\exp-\frac{E_{phys}-E_S}{T_d} & \text{; if } E_S > 0 \end{cases}$$

$$\alpha_{pcD} = \begin{cases} \frac{16\nu_{D_P}T_d}{(E_{chem}-E_S)}\exp\left(-2a_{pc}\sqrt{\frac{4m_Hk(E_{phys}-E_S)}{\hbar^2}}\right) = \frac{16\nu_{D_P}T_d}{(E_{chem}-E_S)}\exp\left(-5.6\times 10^9 a_{pc}\sqrt{(E_{phys}-E_S)}\right) & \text{; if } E_S < 0 \\ 4\nu_{H_P}\left(1+\sqrt{\frac{E_{chem}-E_S}{E_{phys}-E_S}}\right)^{-2}\exp-\frac{E_{phys}-E_S}{T_d} & \text{; if } E_S > 0. \end{cases}$$

The first expressions for $\alpha_{pcH}$ and $\alpha_{pcD}$ show the mobilities to go from a physisorbed site to a chemisorbed site by the tunneling effect. The second expressions represent the mobilities by thermal hopping. The atoms from the gas phase can arrive directly in a chemisorbed site with a probability $T_H$ and $T_D$. These probabilities are also used to calculate the mobilities $\alpha_{pc}$. The mobility is the probability times the oscillation factor $\nu$, and the energy of the atom coming from the gas phase is $E_{phys}+Tg$, as shown in fig 1.

$$T_H \sim T_D = 4\left(1+\sqrt{\frac{E_{chem}-E_S}{E_{phys}-E_S}}\right)^{-2}\exp-\frac{E_{phys}-E_S}{E_{phys}+T_g} \quad (5)$$

The different rates for an H and an D atom to evaporate from a physisorbed site are: $\beta_{Hp} = \nu_{Hp} \exp\frac{-E_{Hp}}{T_d}$ and $\beta_{Dp} = \nu_{Dp} \exp\frac{-E_{Dp}}{T_d}$. We can now derive final expressions for the formation efficiencies of $H_2$ and HD that depend only on the characteristics of the different surfaces. These characteristics are presented in table 1.

For carbonaceous grains (PAHs and amorphous carbon), the formation of $H_2$ and HD is insured by physisorbed atoms populating chemisorbed sites (for temperatures $\geq 20$ K). In this sense, the formation with direct chemisorption from the gas phase is negligible. The term $\frac{T_H}{2}$ can therefore be ignored and the $H_2$ and HD formation efficiencies can be approximated as follows:

$$\epsilon_{H_2}^{carbon} = \epsilon_{HD}^{carbon} = \frac{1 - T_H}{\left(1 + \frac{1}{4}\left(1 + \sqrt{\frac{E_{chem} - E_S}{E_{phys} - E_S}}\right)^2 \exp-\frac{E_S}{T_d}\right)}. \tag{6}$$

For silicate grains, the efficiencies are different due to the fact that chemisorbed sites are populated by the tunnelling effect, and therefore are less easy to access. Also, because $T_H << 1$ due to the high barrier against chemisorption, the efficiencies become:

$$\epsilon_{H_2}^{silicate} = \frac{1}{1 + \frac{16 T_d}{E_{chem} - E_S} \exp-\frac{E_{phys}}{T_d} \exp(4 \times 10^9 a_{pc} \sqrt{E_{phys} - E_S})} + 2\frac{\exp-\frac{E_{phys} - E_S}{E_{phys} + T_g}}{\left(1 + \sqrt{\frac{E_{chem} - E_S}{E_{phys} - E_S}}\right)^2}, \tag{7}$$

$$\epsilon_{HD}^{silicate} = \frac{1}{1 + \frac{16 T_d}{E_{chem} - E_S} \exp-\frac{E_{phys}}{T_d} \exp(5.6 \times 10^9 a_{pc} \sqrt{E_{phys} - E_S})} + 2\frac{\exp-\frac{E_{phys} - E_S}{E_{phys} + T_g}}{\left(1 + \sqrt{\frac{E_{chem} - E_S}{E_{phys} - E_S}}\right)^2}. \tag{8}$$

2.2. Formation rates of $H_2$ and HD on dust surfaces

In astrophysical environments, $H_2$ and HD formation rates (in cm$^{-3}$s$^{-1}$) on dust grains are written as:

$$R_d(H_2) = \frac{1}{2} n(H) v_H n_{grain} \sigma \epsilon_{H_2} S(T_g, T_d), \tag{9}$$

$$R_d(HD) = n(D) v_D n_{grain} \sigma \epsilon_{HD} S(T_g, T_d), \tag{10}$$

where $n(H)$ and $n(D)$ are the number densities of H and D atoms in the gas phase, $v_H$ and $v_D$ are the thermal velocities of H and D atoms calculated as $\sqrt{\frac{8\pi k T_g}{m_H}}$ and $\sqrt{\frac{8\pi k T_g}{m_D}}$, $\epsilon_{H_2}$ and $\epsilon_{HD}$ are the formation efficiencies of $H_2$ and HD respectively. The sticking coefficient, $S(T_g, T_d)$, depends on both the gas temperature $T_g$ and dust grain temperature $T_d$, and is derived by Burke & Hollenbach (1983). This coefficient can be written as: $S(T_g, T_d) = \left(1 + 0.4 \times \left(\frac{T_g + T_d}{100}\right)^{0.5} + 0.2 \times \frac{T_g}{100} + 0.08 \times \left(\frac{T_g}{100}\right)^2\right)^{-1}$. The sticking coefficient is equal to unity for low gas and dust temperatures, and decreases strongly as gas temperature increases. The mean cross section for collisions between grains and atoms per H atom, $\frac{n_{grain}\sigma}{n_H}$, is determined as $< \frac{n_{grain}(a)}{n_H} \pi a^2 >$ with $n_{grain}(a)$ the grain number density for particles with radius between $a$ and $a+da$. In the ISM, this cross section can be estimated for different grain size distributions. Weintgarner & Draine (2001), who consider carbon and silicate grains to model the Milky Way dust, find that very small carbon grains (PAHs) are quite abundant. The total cross section per H atom for the formation of molecules using this distribution is $\frac{n_{grain}\sigma}{n_H} = 2.8\ 10^{-21}$ cm$^{-2}$. The cross section represented by the surface of PAHs (small grains below 100Å) is $1.6\ 10^{-21}$cm$^{-2}$ and the one by amorphous carbon grains is $1.7\ 10^{-22}$ cm$^{-2}$, whereas the surface represented by silicate grains is $10^{-21}$ cm$^{-2}$. With this distribution, it seems clear that small grains (PAHs) constitute most of the cross section available for grain surface chemistry (∼55%). Another grain size distribution, from Mathis, Rumpl & Nordsieck (1977), considers that carbon and silicate grains are almost equally abundant, and also that the distribution starts for grains larger than 50Å and follows a power law slope in size of -3.5. With this distribution, the cross section per H atom is 1.1 $10^{-21}$ cm$^{-2}$. In this case, the carbon and silicate grains represent the same total cross section of ∼5.6 $10^{-22}$ cm$^{-2}$. The formation rates of H$_2$ and HD have to take into account the cross section from carbon and silicate grains as well as the efficiencies on these grains:

$$R_d(H_2) = \frac{1}{2} n(H) v_H S(T_g, T_d) \left( (n_{grain} \sigma \epsilon_{H_2})^{carbon} + (n_{grain} \sigma \epsilon_{H_2})^{silicate} \right) \tag{11}$$

$$R_d(HD) = n(D) v_D S(T_g, T_d) \left( (n_{grain} \sigma \epsilon^{HD})^{carbon} + (n_{grain} \sigma \epsilon_{HD})^{silicate} \right) \tag{12}$$

In the high redshift universe, grain size distributions are different than in our Milky Way, and at lower metallicity, grains are supposed to be smaller. Models of grain size distributions have been made by Todini & Ferrara (2001) for type II SN progenitors with different metallicities. In this study, carbon grains, produced in the ejecta of the supernovae, have a grain size distribution that slightly changes with the metallicity of the progenitor, while silicate grain size distributions are strongly affected by their metallicities. Bianchi & Schneider (2007) show that dust grains produced in the supernova ejecta are also affected by the reverse shock that destroys a big fraction of dust grains and creates smaller grains. The result of the reverse shock is a constant grain size distribution until very small grain sizes, and therefore an increase of the total cross section of dust grains.

Because our goal is to model the formation of HD and H$_2$ at high redshift, we assume that at solar metallicity, the total dust grain cross section is equal to the ISM cross section derived by Weingartner & Draine (2001) of 2.8 $10^{-21}$ cm$^{-2}$, and that the dust abundance scales linearly with the overall metallicity, ignoring corrections due to non-solar elemental abundance ratios. In this sense, by using a linear scale we do not take into account the fact that grains might be smaller at higher redshift, which could increase the total cross section considerably. For exploratory purposes we therefore consider linear scaling, keeping in mind that the cross sections derived as a function of metallicity are lower limits, so that the H$_2$ and HD formation rate can be written as:

$$R_d(H_2) = \frac{7.25 \times 10^{-15} n(H) n_H \sqrt{\frac{T_g}{100}}}{\left( 1 + 0.4 \times \left( \frac{T_g + T_d}{100} \right)^{0.5} + 0.2 \times \frac{T_g}{100} + 0.08 \times \left( \frac{T_g}{100} \right)^2 \right)} \frac{Z}{Z_\odot} (1.75 \epsilon_{H_2}^{carbon} + 1.1 \epsilon_{H_2}^{silicate}), \tag{13}$$

$$R_d(HD) = \frac{1.1 \times 10^{-14} n(D) n_H \sqrt{\frac{T_g}{100}}}{\left( 1 + 0.4 \times \left( \frac{T_g + T_d}{100} \right)^{0.5} + 0.2 \times \frac{T_g}{100} + 0.08 \times \left( \frac{T_g}{100} \right)^2 \right)} \frac{Z}{Z_\odot} (1.75 \epsilon_{HD}^{carbon} + 1.1 \epsilon_{HD}^{silicate}). \tag{14}$$

The more general expressions (11) and (12) should be used to incorporate deviations from the Galactic dust grain size distribution.

## 3. Chemical model and results

### 3.1. Gas phase

Gas phase chemistry of HD in the early Universe has been discussed in great detail by Glover & Abel (2008, and references therein). These authors study the effects of the uncertainties of the different rate coefficients on H$_2$ and HD chemistry and cooling. We here consider only those reactions that are needed to form and destroy $H_2$ and HD, such that the impact of grain surface reactions can be assessed. The different reactions considered in our calculations are summarized in table 1. The formation rates of $H_2$ and HD through the $H^-$ + H route and the $D^-$ + H and $H^-$ + D routes have uncertainties, as discussed by Glover & Abel (2008). We list in the table the low and high values for these rates and consider in our calculations only the high values in order to derive conservative estimates for the contributions of grain surface reactions.

### 3.2. Grain and gas coupling

The gas phase and grain surface chemical models described above are coupled in order to follow the relative importance of grain and gas chemistry in the formation of $H_2$ and HD. The abundance of each species, on grains and in the gas phase, is calculated through a system of rate equations. These equations are solved using the DVODE solver (Brown, Byrne & Hindmarsh 1989) with fixed-leading coefficient implementation. For the grain surface chemistry, we follow the evolution of physisorbed H and D, $H_2$, HD and $D_2$, chemisorbed H and D and for the gas phase chemistry, the evolution of H, $H^+$, $H^-$, D, $D^+$, $D^-$, $H_2$, HD, $D_2$, and $H_2^+$. The formation of $D_2$, as discussed in Cazaux et al. (2008), is never relevant.

In the case of the grain surface chemistry, the incoming fluxes of H and D atoms from the gas phase are in MLyrs/s (monolayers/s). These fluxes are calculated as follows:

$$F_X = \frac{n(X)v_X}{N_S}, \tag{15}$$

where $N_S$ is the number of sites per $cm^2$ on the surface of the grain, $n(X)$ is the density of H or D atoms in the gas phase, and $v_X$ is the mean velocity of these atoms. We assume a density of sites on the grain equal to $N_S \sim 2 \times 10^{15}$ sites $cm^{-2}$, calculated as $\frac{1}{a^2}$, where a is the distance between two (physisorbed or chemisorbed) sites.

To extend the rate equations that describe the chemistry in the gas phase, we convert the densities of the species that are released into the gas phase by dust grains into atoms $cm^{-3}$ $s^{-1}$. This allows us to compare grain surface and gas phase chemistries on the level of rates.

### 3.3. Cloud collapse and thermodynamics

We assume our model clouds to have a uniform metallicity and to undergo a gravitational collapse at the free-fall rate. Metal enrichments of $10^{-5}Z_\odot$, $10^{-4}Z_\odot$ and $10^{-3}Z_\odot$, are considered, motivated by Bromm & Loeb (2003). The simulations start with a density of 1 $cm^{-3}$ at 1,000 K, include dust-grain thermal coupling by collisions and are exposed to a modest cosmic ray ionization rate of $10^{-18}$ $s^{-1}$, appropriate for a situation where prior massive star formation and metal enrichment has occurred. It is assumed that the cloud sees a mean Lyman Werner UV background of 40 in units of $10^{-21}$ erg $s^{-1}$ $cm^{-2}$ $sr^{-1}$ $Hz^{-1}$, modest enough to allow $H_2$ self-shielding (Dijkstra et al. 2008). The helium and deuterium abundances are 0.0825 and $2.6 \times 10^{-5}$, respectively. In general, we follow the set-up as presented in Glover & Savin (2008). We adopt a fiducial redshift of $z = 10$, which yields a temperature of 30 K for the cosmic microwave background (CMB). The dust temperature is set equal to the CMB temperature. The gas density evolution follows:

$$\frac{d\rho}{dt} = \frac{\rho}{t_{ff}} \tag{16}$$

where $t_{ff} = \sqrt{3\pi/32G\rho}$ is the free-fall time. During the collapse, as described in Glover & Abel (2008), the gas temperature follows from the energy equation and evolves as:

$$\frac{dT}{dt} = \frac{\gamma - 1}{\rho}[T\frac{d\rho}{dt} - \frac{\mu}{k}(\Lambda - \Gamma)] + \frac{T}{\gamma - 1}\frac{d\gamma}{dt} + T\frac{dlog\mu}{dt} \tag{17}$$

where $\gamma$ is the adiabatic index, which remains close to 5/3 during the collapse, $\mu$ is the mean molecular weight (generally close to its atomic value), $\Lambda$ is the total cooling rate per unit volume, and $\Gamma$ is the total heating rate per unit volume. The different time derivative terms reflect the usual processes like adiabatic compression and changes in the chemical composition of the gas. The cooling rate includes contributions of $H_2$ and HD as in Glover & Abel (2008) and all fine-structure lines as described in Meijerink & Spaans (2005), with level populations computed under statistical equilibrium. Contributions from $H_2$ formation heating and collisional de-excitation of vibrationally excited $H_2$ have been added. Typical temperature profiles for different metallicities are presented in fig 5. As expected, a medium with a higher metallicity is cooled much more efficiently than a medium with fewer metals. The lowest temperature reached is set by the CMB temperature (at a redshift of 10), and holds for densities that approach and exceed the critical density of the dominant cooling lines (including some neutral carbon emission). At metallicities $\leq 10^{-4} Z_\odot$, the temperature rises at high densities after an initial decrease. This is due to a transition from sub-thermal ($\propto n_H^2$) to thermal ($\propto n_H$) level population excitation of the $H_2$ and HD cooling lines, while gravitational heating scales $\propto n_H^{1.5}$.

### 3.4. Chemical results

The fractional abundances of the different species, as well as the rate of formation of $H_2$ and HD are presented in fig 6. Grain surface chemistry plays an important role in the formation of $H_2$ even for a collapsing cloud with a very low metallicity of $10^{-5} - 10^{-4}$ solar, while the $H^-$ route mostly dominates in diffuse environments, below $10^3$ cm$^{-3}$ (see also Cazaux & Spaans 2004). The formation of HD, on the other hand, is always dominated by gas phase reactions. Indeed, HD forms mostly through the association of $D^+$ and $H_2$. Once the dust grains boost the formation of $H_2$, the $H_2$ abundance increases, and favors the gas phase formation of HD even more. The HD grain surface route scales with metallicity, but also with the amount of neutral hydrogen and deuterium available in the gas phase. Therefore, as the metallicity increases, $H_2$ formation is boosted by dust grains and H is converted into $H_2$. This leads to a higher amount of $H_2$ to form HD through gas phase reactions, and a lower amount of H to form HD through grain surface reactions.

The abundances show that at very low metallicities ($Z = 10^{-5} - 10^{-4} Z_\odot$), neutral hydrogen and deuterium are not completely converted into molecular form when the cloud reaches $n_H = 10^8$ cm$^{-3}$. This result is in agreement with the calculations of Glover & Savin (2008). As the medium contains more metals, the H/$H_2$ front appears before $n_H = 10^8$ cm$^{-3}$, and the D/HD conversion front occurs at earlier stages of the collapse. Once $H_2$ is available in the medium, the $D^+ + H_2$ route is so efficient, despite the dropping $D^+$ abundance, that all deuterium is converted quickly into HD. This D/HD conversion occurs at densities $< 10^3$ cm$^{-3}$ for metallicities $> 10^{-4} Z_\odot$. In this range, the HD level populations are not yet thermalized by collisions. Hence, the associated cooling rates scale $\propto n_H^2$, and help to cool the medium down.

## 4. Conclusions and Discussion

In general, we find that grain surface reactions make a significant contribution to the formation of $H_2$, even at metallicities as low as $Z = 10^{-5} Z_\odot$. It seems that HD formation is driven by gas phase chemistry routes through the association of $D^+$ and $H_2$, helped by the fact that the $H_2$ abundance is strongly boosted by grain surface reactions for all considered metallicities.

In our simulations, we have fixed the grain temperature at the CMB temperature. This might be true if the dust grains are able to cool efficiently irrespective of their size, and if dust-gas coupling is modest. The latter coupling dominates for densities obeying $n_c \sim 10^{4.5}/Z$ cm$^{-3}$, with $Z$ in solar units (Schneider et al. 2006). For the metallicities considered here, $n_c$ is larger than the critical densities of the dominant HD and $H_2$ cooling lines. Hence, the gas is not able to heat up the dust in this case.

However, dust grains in high redshift environments are likely to be in the form of very small graphitic grains (PAHs), which have quite modest heat capacities. Consequently, they can enjoy large excursions in temperature, upto a few hundred Kelvin, when exposed to a soft (non-dissociative UV-visual) background radiation field (Draine & Li 2001). Interestingly, experiments on graphite by Zecho et al. (2002) show that there is an important isotopic effect between the formation of $H_2$ and HD. In their experiments, $H_2$ formation is effective until a lower surface temperature than HD formation. These authors suggest that this effect could be the result of a higher binding energy of deuterium compared to hydrogen on graphite. If we assume these binding energies for the formation of $H_2$ and HD, then we see that the efficiency for the formation of $H_2$ declines more rapidly with increasing dust temperature than the corresponding rate of HD, see fig 7. Therefore, if dust grains are warm ($\sim 150$ K), then $H_2$ formation can be suppressed, while HD formation remains fast. This would lead to a mode where HD is formed more efficiently on grain surfaces than in the gas phase.

## 5. Appendix: Mobility of H and D atoms.

In a previous paper, Cazaux & Tielens (2004) calculated the different mobilities for an atom to go from a site $i$ (physisorbed or chemisorbed) to a site $j$ (physisorbed or chemisorbed). These mobilities depend on the transmission coefficients to cross the barrier as:

$$\alpha_{ij} = \nu_i \times P_{ij} \tag{18}$$

where $\nu_i$ is the oscillation factor in the site $i$, and $P_{ij}$ is the probability for an atom to go from a site $i$ to a site $j$. We consider the atoms thermalized with the dust grain, and therefore that their energies $E$ follow a Boltzmann distribution with mean value equal to the temperature of the dust. The probabilities to go from site to site depend on the transmission coefficients to cross the barriers separating these sites (which we consider to be square). These probabilities are written as:

$$P_{ij} = \frac{1}{kT_d} \int_0^{B_i} exp(-\frac{E}{kT_d}) T_{ij}^{(1)} dE + \frac{1}{kT_d} \int_{B_i}^{\infty} exp(-\frac{E}{kT_d}) T_{ij}^{(2)} dE \tag{19}$$

The first term of this expression shows the probability to cross the barrier through tunneling effects, meaning that the energy of the atom is lower than the barrier $B_i$, while the second term shows the probability through thermal hopping, meaning that the energy of the atom is higher than $B_i$. $T_{ij}^{(1)}$ and $T_{ij}^{(2)}$ are the transmission coefficients for tunneling and thermal hopping, respectively. In this work, we are interested in the mobility to go from a physisorbed site to a chemisorbed site. As shown in Cazaux & Tielens (2004), the transmission coefficients $T_{pc}$ are written as:

$$T_{pc}^{(1)} = 4\left(\left(1 + \sqrt{\frac{E-(E_{phys}-E_{chem})}{E}}\right)^2 + \frac{(E_{phys}-E_S)(E_{chem}-E_S)(\sinh\sqrt{\frac{2mk(E_{phys}-E_S-E)}{\hbar^2}}a_{pc})^2}{(E_{phys}-E_S-E)E}\right)^{-1} \;;if E < (E_{phys}-E_S) \quad (20)$$

$$T_{pc}^{(2)} = 4\left(\left(1 + \sqrt{\frac{E-(E_{phys}-E_{chem})}{E}}\right)^2 - \frac{(E_{phys}-E_S)(E_{chem}-E_S)(\sin\sqrt{\frac{2mk(E-(E_{phys}-E_S))}{\hbar^2}}a_{pc})^2}{(E_{phys}-E_S-E)E}\right)^{-1} \;;if E > (E_{phys}-E_S) \quad (21)$$

The probabilities and mobilities of H and D atoms are then calculated using equations (18) and (19). Because of the weight of the Boltzmann distribution, the energies of the atoms that tunnel are mostly very small, of the order of the temperature of the dust. Therefore, only the second term of equation (20) is important. For thermal hopping, on the other hand, the energies of the atoms are higher than the energy of the barrier, and the first term of equation (20) dominates. In this way, we can derive the approximations used in this paper.

*Acknowledgements.* We would like to thank the referee, Paola Caselli, for very constructive comments and careful reading of the manuscript that improved considerably the quality of this paper.

**References**

Abel, T., Bryan, G. L., & Norman, M. L. 2000, ApJ, 540, 39

Abel, T., Wise, J. H., & Bryan, G. L. 2007, ApJ, 659, L87

Bachellerie, D., Sizun, M., Teillet-Billy, D., Rougeau, N. and Sidis,V. Chem. Phys. Let. 448, (2007) p 223

Bianchi, S., & Schneider, R. 2007, MNRAS, 378, 973

Bromm, V., Coppi, P. S., & Larson, R. B. 2002, ApJ, 564, 23

Bromm, V., & Loeb, A. 2003, Nature, 425, 812

Brown, P.N., Byrne, G.D., & Hindmarsh, A.C., 1989, SIAM J. Sci. Stat. Comput., 10, 1038

Burke, J. R., & Hollenbach, D. J. 1983, ApJ, 265, 223

Cazaux, S., & Tielens, A. G. G. M. 2002, ApJ, 575, L29

Cazaux, S., & Spaans, M. 2004, ApJ, 611, 40

Cazaux, S., Caselli, P., Cobut, V., & Le Bourlot, J. 2008, A&A, 483, 495

Dijkstra, M., Haiman, Z., Mesinger, A., & Wyithe, J. S. B. 2008, MNRAS, 391, 1961

Draine, B.T. & Li, A., 2001, ApJ, 551, 807

Galli, D., & Palla, F. 1998, A&A, 335, 403

Glover, S. C. O., & Savin, D. W. 2008, arXiv:0809.0780

Glover, S. C. O., & Abel, T. 2008, ArXiv e-prints, 803, arXiv:0803.1768

Glover, S. C. O., & Jappsen, A.-K. 2007, ApJ, 666, 1

Hornekær, L., Sljivancanin, Z., Xu, W., Otero, R., Rauls, E., Stensgaard, I., Lægsgaard, E., Hammer, B., Besenbacher, F. 2006,, Phys. Rev. Lett., vol. 96, pp. 156104.

Jappsen, A.-K., Glover, S. C. O., Klessen, R. S., & Mac Low, M.-M. 2007, ApJ, 660, 1332

Jeloaica, L. & Sidis, V., Chem. Phys. Lett., Volume 300, Issues 1-2, 29 January 1999, Pages 157-162

Johnson, J. L., & Bromm, V. 2006, MNRAS, 366, 247

Klessen, R. S., Spaans, M., & Jappsen, A.-K. 2007, MNRAS, 374, L29

Maiolino, R., Schneider, R., Oliva, E., Bianchi, S., Ferrara, A., Mannucci, F., Pedani, M., & Roca Sogorb, M. 2004, Nature, 431, 533

Mathis, J. S., Rumpl, W., & Nordsieck, K. H. 1977, ApJ, 217, 425

McGreer, I. D., & Bryan, G. L. 2008, ApJ, 685, 8 e-prints, 802, arXiv:0802.3918

Meijerink, R. & Spaans, M., 2005, A&A, 436, 397

Nozawa, T., Kozasa, T., Umeda, H., Maeda, K., & Nomoto, K. 2003, ApJ, 598, 785

Omukai, K., Tsuribe, T., Schneider, R., & Ferrara, A. 2005, ApJ, 626, 627

Ripamonti, E. 2007, MNRAS, 376, 709

Rougeau,N., Teillet-Billy, D. and Sidis, V., Chem. Phys. Lett. 431 (2006), p. 135

Sha, X., & Jackson B., Surface Science, Volume 496, Issue 3, 10 January 2002, Pages 318-330

Smith, B., Sigurdsson, S., & Abel, T. 2008, MNRAS, 385, 1443

Schneider, R., Omukai, K., Inoue, A.K., & Ferrara, A., 2006, MNRAS, 369, 1437

Spaans, M., & Silk, J. 2005, ApJ, 626, 644

Todini, P., & Ferrara, A. 2001, MNRAS, 325, 726

Walmsley, C. M., Flower, D. R., & Pineau des Forêts, G. 2004, A&A, 418, 1035

Weingartner,J. C., & Draine, B. T. 2001, ApJ, 548, 296

Wise, J. H., & Abel, T. 2007, ApJ, 671, 1559

Wise, J. H., & Abel, T. 2008, ApJ, 684, 1

Yoshida, N., Omukai, K., Hernquist, L., & Abel, T. 2006, ApJ, 652, 6

Yoshida, N., Omukai, K., & Hernquist, L. 2007, ApJ, 667, L117

Zecho, T., Guttler, A., Sha, X., Jackson, B., & Kuppers, J. 2002, J. Chem. Phys., 117, 8486

**Table 1.**

| Surface | $E_{phys}$ | $E_{Hp}$ | $E_{Dp}$ | $E_{chem}$ | $E_S$ | $a_{pc}$ | $\nu_{Hp}$ | $\nu_{Dp}$ |
|---|---|---|---|---|---|---|---|---|
| Graphite (PAHs) | 800 K | 720K | 745K | 7000 K | -2300 K | 1.5 Å | $3\ 10^{12}$ | $2\ 10^{12}$ |
| Para sites (PAHs) | 800 K | 720K | 745K | 25000 K | 200 K | 1.5 Å | $3\ 10^{12}$ | $2\ 10^{12}$ |
| Amorphous Carbon | 800 K | 720K | 745K | 7000 K | 200 K | 3 Å | $3\ 10^{12}$ | $2\ 10^{12}$ |
| Silicates | 700 K | 630K | 650K | 15000 K | -1000 K | 1.7Å | $3\ 10^{12}$ | $2\ 10^{12}$ |

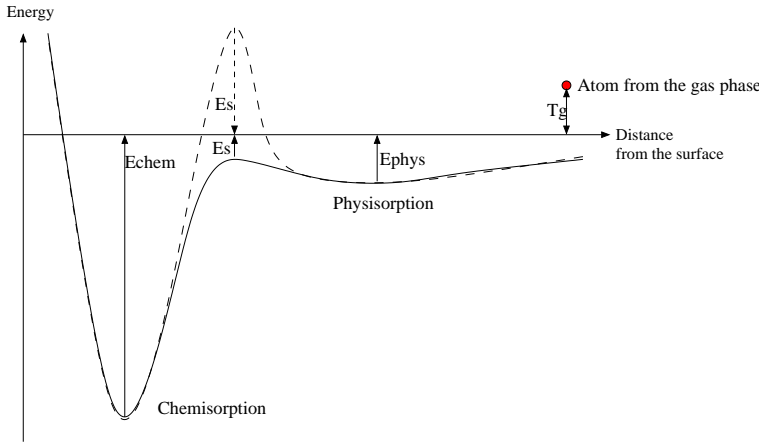

**Fig. 1.** Interactions between an atom and a surface. Two types of surfaces are represented, the one with a high barrier against chemisorption (where $E_S \leq 0$) and the one with no barrier against chemisorption (with $E_S \geq 0$). Note that an atom coming from the gas phase has an energy $E_{phys} + T_g$.

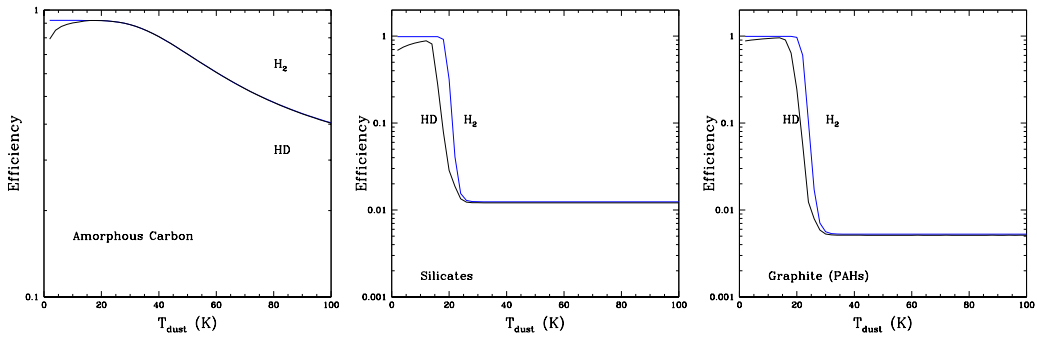

**Fig. 2.** $H_2$ and HD formation efficiencies on amorphous carbon (left), silicates (middle) and PAHs (right) as a function of the grain temperature. These efficiencies have been calculated using a rate equations method.

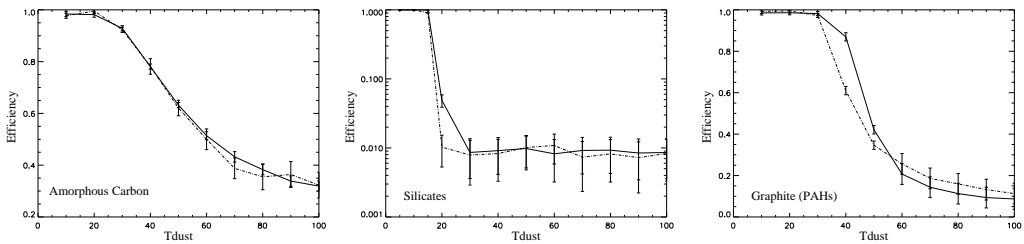

**Fig. 3.** $H_2$ and HD formation efficiencies on amorphous carbon (left), silicates (middle) and PAHs (right) as a function of the grain temperature. These efficiencies have been calculated using Monte Carlo simulations. For the case of PAHs, the properties of para-sites have been included, resulting in a much higher efficiency.

**Table 2.** Reactions and rate coefficients adopted in the chemical model.

| Cosmic rays ionization | rate in s$^{-1}$, with $\zeta$ the rate of cosmic ray ionization of H$_2$. | ref |
|---|---|---|
| H + CR → H+ + e | 0.46$\zeta$ | c |
| D + CR → H+ + e | 0.46$\zeta$ | c |
| H2 + CR → H + H | 1.5$\zeta$ | c |
| H2 + CR → H2+ + e | 0.96$\zeta$ | c |
| HD + CR → H + D | 1.5$\zeta$ | c |
| HD + CR → HD+ + e | 0.96$\zeta$ | c |
| D2 + CR → D + D | 1.5$\zeta$ | c |

| Reaction | rate in cm$^3$ s$^{-1}$ | ref |
|---|---|---|
| H2 + D → H + HD | $dex(-56.4737 + 5.88886 \times log(Tg) + 7.196292 \times log(Tg)^2 + 2.25069 \times log(Tg)^3 - 2.16903 \times log(Tg)^4$ $+ 0.317887 \times log(Tg)^5$ | a |
| HD + H → H2 + D | $5.25 \times 10^{-11} \times \exp(-4430/Tg)$ if Tg < 200K $5.25 \times 10^{-11} \times \exp((-4430/Tg) + (173900/Tg^2))$ if Tg > 200K | a |
| D+ + H2 → H+ + HD | $10^{-9} \times (0.417 + 0.846 \times log(Tg) - 0.137 \times log(Tg)^2)$ | a |
| H+ + HD → D+ + H2 | $1.1 \times 10^{-9} \times \exp -488/Tg$ | a |
| H+ + D → D+ + H | $2 \times 10^{-10} \times Tg^{0.402} \times \exp -37.1/Tg - 3.31 \times 10^{-17} \times Tg^{1.48}$ | a |
| H + D+ → H+ + D | $2.06 \times 10^{-10} \times Tg^{0.396} \times \exp -33/Tg + 2.03 \times 10^{-9} \times Tg^{-0.332}$ | a |
| HD + D+ → D2 + H+ | $1 \times 10^{-9}$ | a |
| H+ + D2 → D+ + HD | $2.1 \times 10^{-9} \times \exp -491/Tg$ | a |
| H2 + H+ → H2+ + H | $\exp(-21237.15/Tg) \times (3.3232183 \times 10^{-7} + 3.3735382 \times 10^{-7} \times ln(Tg) - 1.4491368 \times 10^{-7} \times ln(Tg)^2$ $+ 3.4172805 \times 10^{-8} \times ln(Tg)^3 - 4.7813720 \times 10^{-9} \times ln(Tg)^4 + 3.9731542 \times 10^{-10} \times ln(Tg)^5$ $- 1.8171411 \times 10^{-11} \times ln(Tg)^6 + 3.5311932 \times 10^{-13} \times ln(Tg)^7$ | b |
| H + H+ → H2+ + phot | $dex(-19.38 - 1.523 \times log(Tg) + 1.118 \times (log(Tg))^2 - 0.1269 \times (log(Tg))^3))$ | a |
| H + H2+ → H2 + H+ | $6.4 \times 10^{-10}$ | a |
| H + HD+ → H2 + D+ | $1 \times 10^{-9}$ | a |
| H+ + e → H + phot | $2.753 \times 10^{-14} \times (315614/Tg)^{1.5} \times (1 + (115188/Tg)^{0.407})^{-2.242}$ | b, caseB |
| D+ + e → D + phot | $2.753 \times 10^{-14} \times (315614/Tg)^{1.5} \times (1 + (115188/Tg)^{0.407})^{-2.242}$ | b, case B |
| H2+ + e → H + H | $10^{-8}$ if Tg < 617K $1.32^{-6} \times Tg^{-0.76}$ if Tg > 617K | a |
| HD+ + e → H + D | $7.2 \times 10^{-8} \times Tg^{-0.5}$ | a |
| H + e → H- + phot | $dex(-17.845 + 0.762 \times log(Tg + 0.1523 \times (log(Tg))^2 - 0.03274 \times log(Tg)^3)$ | a |
| D + e → D- + phot | $dex(-17.845 + 0.762 \times log(Tg) + 0.1523 \times (log(Tg))^2 - 0.03274 \times log(Tg)^3)$ | a |
| H + e → H+ + e + e | $\exp(-3.271396 \times 10 + 1.3536 \times 10 \times ln(Te) - 5.7393 \times ln(Te)^2 + 1.5631 \times ln(Te)^3)$ | a |
| D + e → D+ + e + e | $\exp(-3.271396 \times 10 + 1.3536 \times 10 \times ln(Te) - 5.7393 \times ln(Te)^2 + 1.5631 \times ln(Te)^3)$ | a |
| H- + H → H2 + e (higher value) | $5 \times 10^{-9}$ | b |
| H- + H → H2 + e (lower value) | $0.65 \times 10^{-9}$ | b |
| D- + H → HD + e (higher value) | $0.5 \times 5 \times 10^{-9}$ | b |
| D- + H → HD + e (lower value) | $0.5 \times 0.65 \times 10^{-9}$ | b |
| H- + D → HD + e (higher value) | $0.5 \times 5 \times 10^{-9}$ | b |
| H- + D → HD + e (lower value) | $0.5 \times 0.65 \times 10^{-9}$ | b |
| D- + D → D2 + e (higher value) | $5 \times 10^{-9}$ | b |
| D- + D → D2 + e (lower value) | $0.65 \times 10^{-9}$ | b |
| H+ + D- → HD+ + e | $1.1 \times 10^{9} \times (Tg/300)^{-0.4}$ | a |
| D+ + H- → HD+ + e | $1.1 \times 10^{9} \times (Tg/300)^{-0.4}$ | a |

[a] Glover & Savin (2008), [b] Abel & Glover (2008), [c] Walmsley et al. 2004

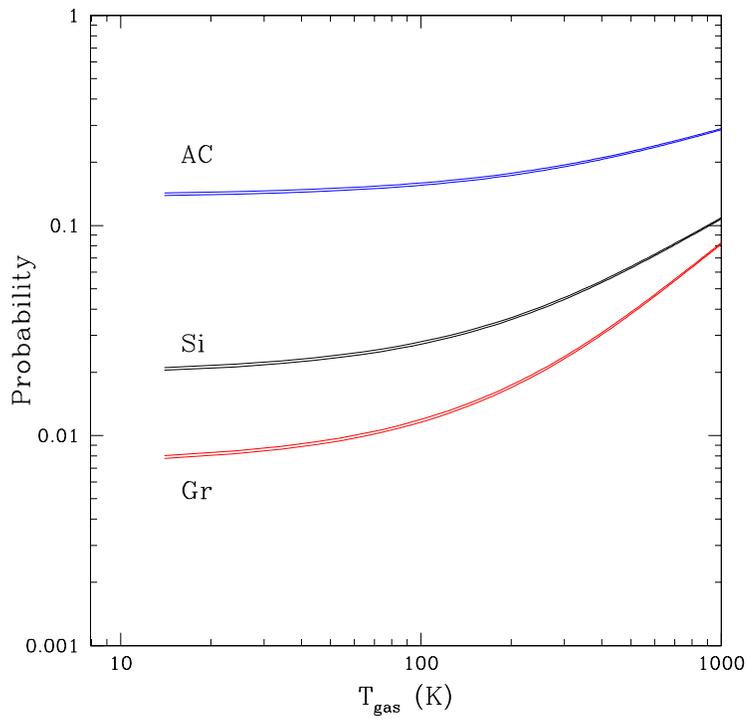

**Fig. 4.** Probability for an atom in the gas phase with a temperature $T_g$ to enter directly into a chemisorbed site on amorphous carbon (AC), silicates (Si) and PAHs (Graphite: Gr).

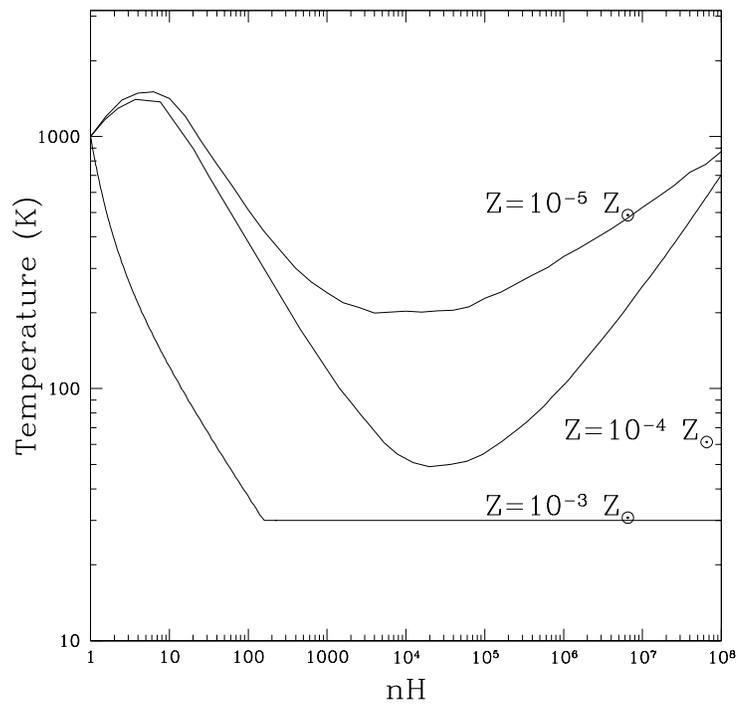

**Fig. 5.** Typical temperature profiles of collapsing gas clouds with different metallicities.

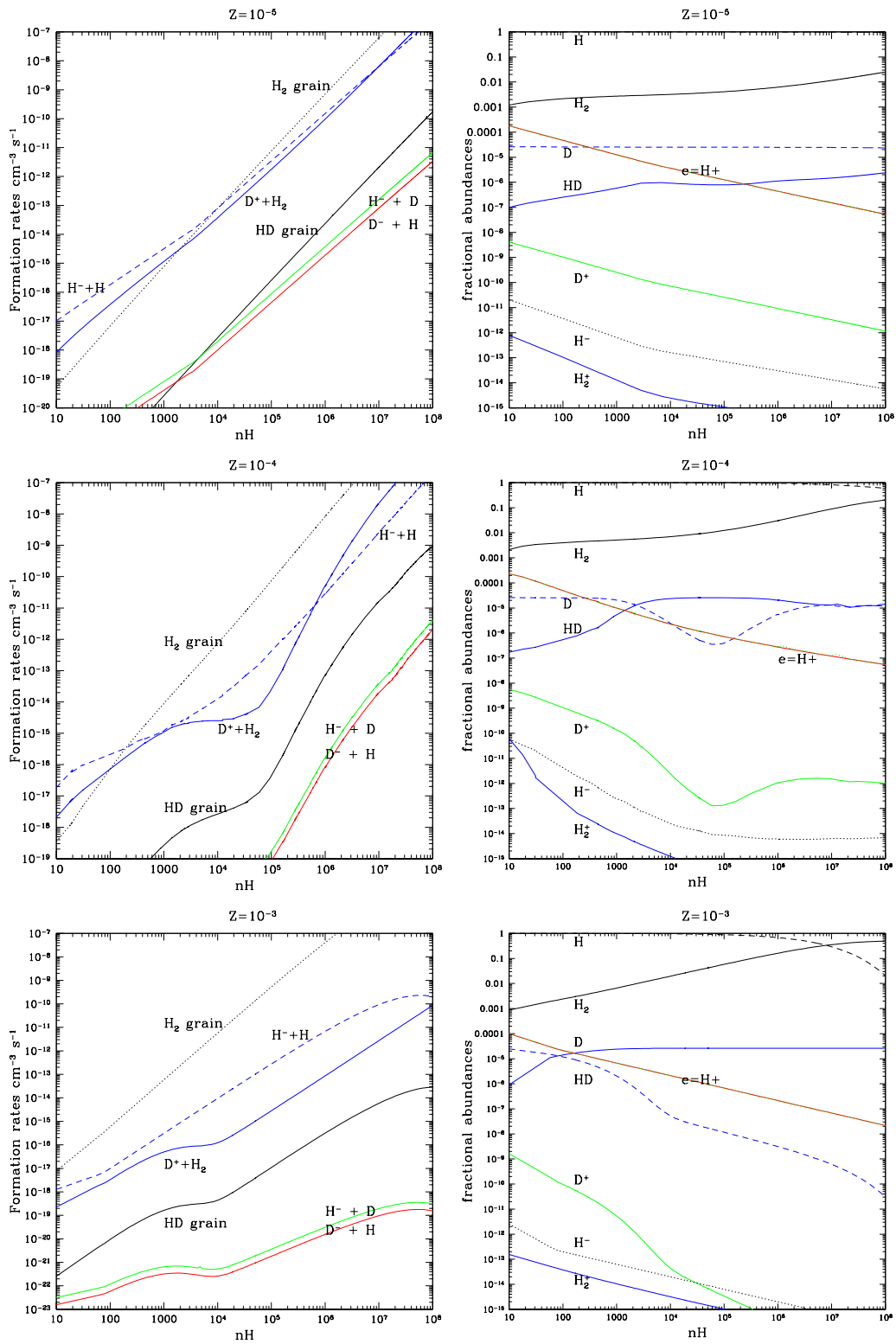

**Fig. 6.** Left panels: formation rates of $H_2$ (dotted and dashed lines) and HD (solid lines) via dust grain and gas phase routes. Right panels: fractional abundances of the species present in the collapsing could. Different metallicities are considered: $Z = 10^{-5} Z_\odot$ (top), $Z = 10^{-4} Z_\odot$ (middle) and $Z = 10^{-3} Z_\odot$ (bottom).

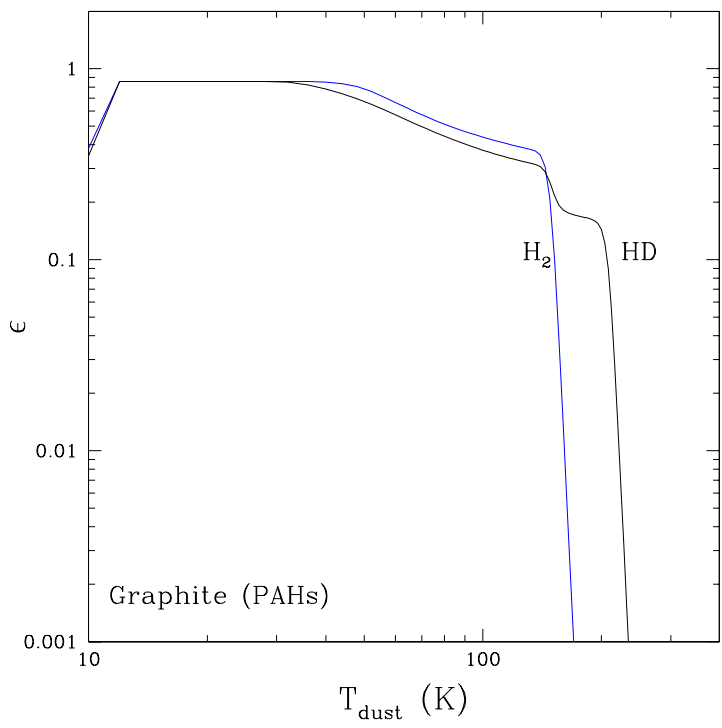

**Fig. 7.** H$_2$ and HD formation efficiencies on graphite (PAHs) surfaces in the high dust temperature limit. The impact of the higher binding energy of D is clearly visible.